\documentstyle[prl,aps,eqsecnum,epsfig,multicol]{revtex}
\begin{document}
\draft

%

\title{Suppression of Hadrons with Large Transverse Momentum 
in Central Au+Au Collisions at $\sqrt{s_{_{NN}}}$~=~130~GeV }

%

\author{
K.~Adcox,$^{40}$
S.{\,}S.~Adler,$^{3}$
N.{\,}N.~Ajitanand,$^{27}$
Y.~Akiba,$^{14}$
J.~Alexander,$^{27}$
L.~Aphecetche,$^{34}$
Y.~Arai,$^{14}$
S.{\,}
% ...
\\(PHENIX Collaboration)\\
}
\address{
% ...
$^{3}$Brookhaven National Laboratory, Upton, NY 11973-5000, USA\\
% ...
$^{14}$KEK, High Energy Accelerator Research Organization, Tsukuba-shi, Ibaraki-ken 305-0801, Japan\\
% ...
$^{27}$Chemistry Department, State University of New York - Stony Brook, Stony Brook, NY 11794, USA\\
% ...
$^{34}$SUBATECH (Ecole des Mines de Nantes, IN2P3/CNRS, Universite de Nantes) BP 20722 - 44307, Nantes-cedex 3, France\\
% ...
$^{40}$Vanderbilt University, Nashville, TN 37235, USA\\
% ...
}

\date{\today}        
\maketitle

%

\begin{abstract}
Transverse momentum spectra for charged hadrons and for 
neutral pions in the range 1~GeV/c~$< p_T <$~5~GeV/c have
been measured by the PHENIX experiment at RHIC in Au+Au 
collisions at $\sqrt{s_{_{NN}}}=130$~GeV. At high $p_T$
the spectra from peripheral nuclear collisions are consistent 
with scaling the spectra from p+p 
collisions by the average number of binary nucleon-nucleon 
collisions.  The spectra from central collisions 
are significantly suppressed when compared to the binary-scaled 
p+p expectation, and also when compared to similarly 
binary-scaled peripheral collisions, indicating a novel nuclear 
medium effect in central nuclear collisions at RHIC energies.
\end{abstract}

\begin{multicols}{2}   
\narrowtext            
%
%
%
%
%
%
%
%

Ultrarelativistic heavy ion collisions provide the opportunity
to study strongly interacting matter at high temperature and density.  
At Brookhaven National Laboratory's Relativistic Heavy Ion Collider (RHIC), 
nuclei as heavy as gold (Au) are accelerated to energies of 
$\sqrt{s_{_{NN}}}=200$~GeV per 
nucleon-nucleon pair.  In the early stages of a central collision, 
energy densities are expected to be sufficient to  
dissolve normal nuclear matter into a phase of deconfined 
quarks and gluons, the ``Quark Gluon Plasma'' (QGP).
The PHENIX experiment is designed to investigate nuclear collisions with
a wide variety of probes, focusing primarily on those produced in 
the early stages of the collision. 

Of particular interest are the products of parton scatterings with
large momentum transfer (``hard scatterings''). 
In p+p collisions hard-scattered partons fragment into jets 
of hadrons; these fragments are the primary source of hadrons at 
high transverse momentum ($p_T$), typically 
above $\sim$2~GeV/c~\cite{textbook}. 
In a high-energy nuclear collision 
hard scattering will occur at the earliest time during the collision, 
well before the QGP is expected to form, and thus the scattered partons 
will subsequently experience the strongly interacting medium created in 
the collision.
These partons are expected to lose energy~\cite{quenching_theory} 
in hot and dense nuclear matter through gluon 
bremsstrahlung, effectively quenching jet production. This would have 
many observable consequences, of which the most directly measurable 
would be a depletion in the yield of high $p_T$ hadrons~\cite{quench_effect}. 
It has been suggested that the energy loss is larger in a medium of deconfined 
color charges than in hadronic matter~\cite{Baier}, making ``jet quenching'' a 
potential signature for QGP formation.   

To quantify such modifications we need a baseline expectation for spectra 
from nuclear (A+A) collisions in the absence of nuclear medium effects. 
Given that hard parton scatterings 
%
%
have small cross sections, 
one can regard the nuclei as an incoherent superposition of partons 
(``point-like scaling''). We approximate this by modelling the A+A 
collision as a sum of independent nucleon-nucleon (N+N) collisions 
(``binary scaling'').  
For a given class of A+A collisions, we can determine 
$\langle N_{binary} \rangle$ the average number of inelastic 
N+N collisions per event and then define the 
{\em nuclear modification factor} as the ratio,

\begin{equation}
R_{AA}(p_T) 
= \frac{(1/N_{evt}) \; d^{2}N^{A+A}/dp_T d\eta }
{(\langle N_{binary} \rangle/\sigma^{N+N}_{inel}) \; d^{2}\sigma^{N+N}/dp_T d\eta}.
\label{eq:RAA_defined}
\end{equation}


%

\noindent
In the absence of nuclear modifications to hard scattering, 
the ratio $R_{AA}$ will be unity; thus departures from $R_{AA}=1$ 
indicate nuclear medium effects. 
%
%
Previous measurements indicate that for $p_{T}$ below 2 GeV/c, $R_{AA}$ is  
smaller than one since the bulk of particle production scales with the 
number of nucleons participating in the reaction~\cite{mult,STAR}.  
%
%
For $p_{T}$ above 2 GeV/c particle production 
in p+A collisions is enhanced compared to binary scaling, 
commonly referred to as the ``Cronin effect''~\cite{cronin}.  
Parton shadowing as measured in lepton+A collisions~\cite{EMC} is
also expected to modify the hadron spectra in p+A and
A+A compared to binary scaling.

%

We examine high-$p_T$ spectra of charged hadrons and neutral pions 
measured by the PHENIX experiment~\cite{daveQM97,billQM01}
in a central and a peripheral class of Au+Au collisions at 
$\sqrt{s_{_{NN}}}=130$~GeV.  
These data are obtained with the central 
spectrometer, which consists of two arms, ``east'' and ``west'', 
each covering $\Delta \phi = 90^{o}$ and $| \eta | <0.35$. The arms 
are positioned outside an axially-symmetric magnetic field 
centered around the beam axis. 

Charged particles are reconstructed using a drift chamber (DC) 
and two layers of multi-wire proportional chambers with pad-readout 
(PC1, PC3) in the east arm. The DC measures the particle 
trajectories between 2.0 and 2.4~m radius in the plane 
perpendicular to the beam axis. A matching hit in PC1 at $\sim2.5$~m,
together with location of the collision vertex, 
fixes the polar angle. 
Particle momenta are determined with a resolution of
$\delta p/p \simeq 0.6\%  \oplus 3.6\%\ p$~(GeV/c). 
The absolute momentum scale is known to better than 2\%. 
Trajectories are confirmed by requiring a matching 
hit within a $\pm 2\sigma$ window (about 2.5~cm) in PC3 at a radius 
of 5~m, which eliminates nearly all secondary tracks 
from decays and interactions in material. The remaining background 
is due primarily to accidental associations.
The level of background to signal is negligible 
below $p_T<2$~GeV/c, rises to 1/10 at 3.5~GeV/c, and reaches 
1/1 at 6~GeV/c.  This background is measured statistically, by 
swapping the z-coordinate of the PC3 hits, and subtracted from 
the yield. 

Corrections for acceptance, reconstruction efficiency, 
decays in flight, momentum resolution, and dead areas are determined 
using a full GEANT simulation.  Simulated single particles are 
embedded in real events to model the effect of 
detector occupancy.  In peripheral Au+Au collisions the track 
reconstruction efficiency exceeds 98\%, while it 
is reduced to $68 \pm 6\%$ for central collisions, independent of momentum.
Corrections due to finite momentum resolution
are negligible at low $p_T$ and rise to the level of 30\% at 5~GeV/c. 
The systematic errors (Table~\ref{tab:syst}) are dominated by the 
uncertainty in the Monte Carlo description of the
detector, including the dead areas and the momentum resolution.


%
%

%
%

\begin{figure}
\centerline{\epsfig{file=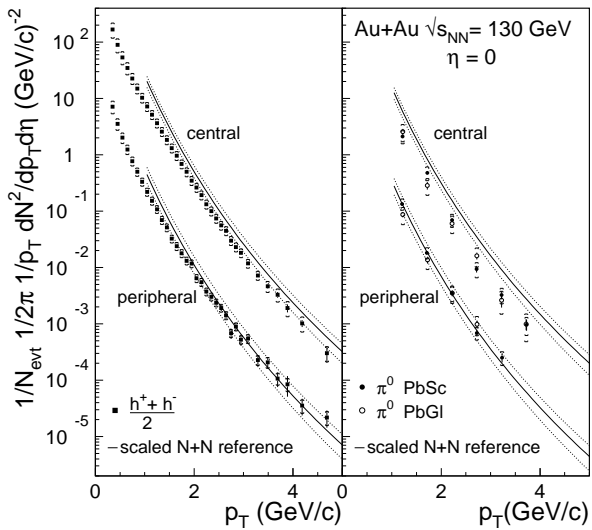,width=1.0\linewidth}}
\caption[]{
	The yields per event at mid-rapidity for 
	charged hadrons ({\em left}) and neutral pions ({\em right}) 
	are shown as a function of $p_T$ for 60--80\%~({\it lower}) 
	and 0--10\%~({\it upper}) event samples.
	The error bars indicate the statistical errors on
	the yield; the surrounding brackets indicate the
	systematic errors.  
	Also shown are the N+N references
	for charged hadrons and neutral pions,
	each scaled up by $\langle N_{binary} \rangle$ for the 
	class.  The bands indicate
	the uncertainty in the N+N reference and in the 
	$\langle N_{binary} \rangle$. }
\label{fig:pt_spectra} 
\end{figure}

Neutral pions are measured via their 
$\pi^0 \rightarrow \gamma \gamma$ decay. Two separate analyses 
are performed, the first using a lead-scintillator (PbSc) sampling 
calorimeter in half ($\Delta \phi = 45^{o}$) of the west arm 
aperture and the second with a lead-glass Cerenkov (PbGl) calorimeter 
in a quarter ($\Delta \phi = 22.5^{o}$) of the east arm aperture.
The two analyses have very different systematics, and 
Figure~\ref{fig:pt_spectra} shows the agreement of their 
final $\pi^{0}$ spectra.

In both analyses, pairs of calorimeter showers are  binned in 
pair $p_T$ and invariant mass $m_{\gamma \gamma}$. 
The energy scale is verified using both the $\pi^{0}$ mass and 
$E/p$ ratio for identified electrons and is 
known to $<1.5$\%~\cite{et}.  Hadron-induced showers are 
suppressed with arrival time and shower shape cuts. 
The combinatorial pair background is estimated by mixing showers 
from different events with similar centrality.  
The mixed $m_{\gamma \gamma}$ distribution is subtracted
from the true distribution after being normalized in a region 
outside the $\pi^0$ mass peak.  The $\pi^{0}$ yield in
each $p_{T}$ bin is determined by integrating the 
subtracted $m_{\gamma \gamma}$ distribution in a window
determined by a Gaussian fit to the $\pi^{0}$ peak.

The $\pi^0$ spectra are corrected for losses
due to energy resolution, cluster overlaps, analysis cuts, 
dead detector areas, and acceptance.  
%
%
Smearing of the photon energy due to resolution and 
cluster overlaps is used to simulate the $m_{\gamma \gamma}$
peak at each $p_{T}$, which agrees with that seen in the data.
To estimate the $\pi^0$ reconstruction efficiency,
the same cuts are applied to the simulated 
$m_{\gamma \gamma}$ distributions as applied to the real data.  

Contributions to the yield from pions not originating from the 
vertex are estimated by GEANT simulation to be 6-8\%.  
The dominant sources of error are the uncertainty in the particle 
identification, the effect of energy smearing, and the peak 
extraction procedure.  Their relative contributions to the total 
errors (see Table~\ref{tab:syst}) differ for the two analyses.

%

Event classification is provided by the combination 
of two beam-beam counters (BBC's) and two zero-degree calorimeters 
(ZDC's).  We present data from two event samples,
central and peripheral. The central sample covers the 0--10\% most 
central fraction of the geometrical Au+Au cross-section, while
the peripheral sample contains events in the 60--80\% selection.
Using a Glauber model combined 
with a simulation of the BBC and ZDC responses~\cite{mult}, we estimate 
$\langle N_{binary} \rangle = 905 \pm 96$ for the central sample,
$\langle N_{binary} \rangle = 20 \pm 6$ for the 
peripheral sample, and $45 \pm 13$ for the ratio between them. 
The errors include the uncertainties in the 
parameters used in the Glauber model~\cite{foot1},
as well as in the fraction of the total geometrical cross section
($92\pm 4\%$) seen by the interaction trigger~\cite{foot2}.

%
%
%

The $p_T$ distributions for charged hadrons and neutral pions 
are shown for both centrality classes in Fig.~\ref{fig:pt_spectra}. 
In this figure, those following, and Table~\ref{tab:syst}, 
the systematic errors shown are the quadrature sums of conservatively 
estimated limits on several independent errors and 
represent upper bounds on standard deviations.
%
%
They are also substantially correlated between points.

The data are compared to the binary-scaled yield from N+N
collisions.  Since no N+N data exist at $\sqrt{s}=$130~GeV,
we parameterize the cross section 
$1/(2 \pi p_{T}) d^2\sigma/d\eta dp_{T}$ for $(h^{+}+h^{-})/2$ as
$A/(1+ p_T/p_0)^n$.  We determine the parameters 
$A$=330~mb/GeV$^2$/c$^2$, $p_{0}$=1.72~GeV/c, and n=12.4
by interpolating between results from p+p collisions at 
the ISR~\cite{alper} and $\overline{\rm p}$+p collisions 
at the Sp$\overline{\rm p}$S~\cite{ua1} and the Tevatron~\cite{cdf}. 
The systematic error in the N+N reference (Table~\ref{tab:syst})
is due to the error in the absolute normalization of the data
used and in the interpolation technique.
For neutral pions we 
scale the charged hadron cross section by the 
charged pion to charged hadron ratio $\pi/h$ observed at the 
ISR~\cite{alper}.  This was found to be $0.63\pm 0.06$ nearly 
independent of $p_T$ above 1.5~GeV/c. 

\begin{figure}
\centerline{\epsfig{file=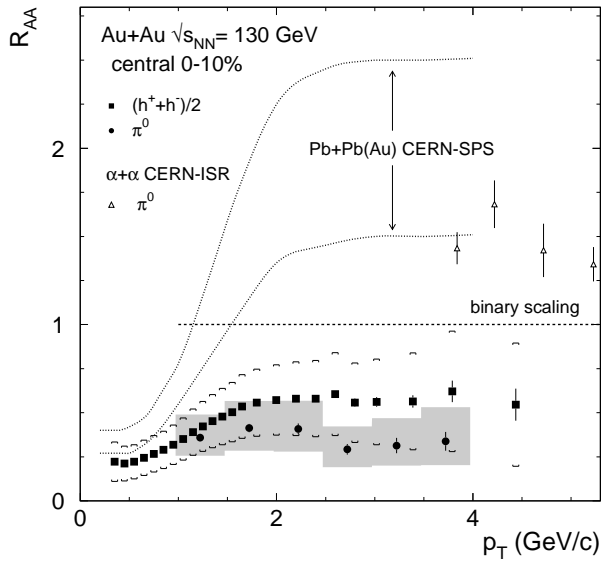,width=1.0\linewidth}}
\caption[]{
	The ratio $R_{AA}$ for charged hadrons
	and neutral pions (weighted average of PbSc and PbGl
	results) in central Au+Au collisions.
	The error bars indicate the statistical errors on the
	measurement.  The surrounding bands [shaded for
	$\pi^{0}$'s, brackets for $(h^{+}+h^{-})/2$]
	are the quadrature sums of (i) the systematic errors on 
	the measurement, 
	(ii) the uncertainty in the N+N reference,
	and (iii) the uncertainty in $\langle N_{binary} \rangle$.
	Also shown are the ratio of 
	inclusive cross sections in $\alpha+\alpha$
	compared to p+p at $\sqrt{s_{_{NN}}}=31$~GeV~\cite{ISR_alphaalpha}, 
	and spectra from central Pb+Pb, Pb+Au compared to p+p
	collisions at $\sqrt{s_{_{NN}}}=17$~GeV~\cite{predict_nuclear_mod}
	shown as a band indicating the range of uncertainty.
	}
\label{fig:ratio_central_pp} 
\end{figure}

For $p_T>2$~GeV/c the binary-scaling prediction
agrees with the data from peripheral collisions
for both charged and neutrals, while for central collisions
the data lie noticeably below the prediction.
To examine this difference more 
directly, we plot the ratio $R_{AA}$ for central collisions 
in Fig.~\ref{fig:ratio_central_pp}.  
For the charged spectrum $R_{AA}$ rises up to 2~GeV/c, as expected;
but above 2~GeV/c $R_{AA}$ remains significantly below unity for 
both spectra.  

The depletion is quite striking, since the production of high-$p_T$ 
hadrons in p+A collisions at fixed-target energies 
is known to be enhanced compared to the binary-scaling expectation 
for $p_T>2$~GeV/c, i.e., the Cronin effect~\cite{cronin}. 
A similar enhancement has also been observed in heavy ion collisions 
at lower energies~\cite{predict_nuclear_mod,ISR_alphaalpha}, 
as shown in Fig.~\ref{fig:ratio_central_pp}.
Phenomenological calculations~\cite{predict_nuclear_mod}  
including shadowing and the Cronin effect predicted that
for central Au+Au collisions at $\sqrt{s_{_{NN}}}$=130~GeV,
$R_{AA} > 1$ for hadron spectra in the $p_T$ range 3--9~GeV/c 
with a peak value of $R_{AA} \simeq 1.3$ at 4~GeV/c.

%

Above 2 GeV/c $R_{AA}$ is lower for pions than for charged 
hadrons, which implies that the $\pi/h$ ratio is smaller in central 
RHIC Au+Au collisions than in ISR p+p collisions.  This is 
consistent with identified charged hadron spectra measured 
by PHENIX~\cite{juliaQM01} for which a large yield of protons 
and antiprotons is observed at $p_T\sim 2$~GeV/c.  


We can also examine the spectra from central collisions for 
modifications at high $p_T$ by comparing
them to the spectra from peripheral collisions after dividing
each by the corresponding values of $\langle N_{binary} \rangle$.
The central-to-peripheral ratio is a useful 
complement to $R_{AA}$, since it should
be unity in the limit of point-like scaling.
Many of the experimental uncertainties
are reduced in this ratio (see Table~\ref{tab:syst}). 
%
%
Additionally, the uncertainty induced by the p+p interpolation
is eliminated, albeit at the expense of incurring that in 
$\langle N_{binary} \rangle$ for the peripheral class.
%
%
We note that there may be effects from the centrality dependence of nuclear
shadowing and/or the Cronin effect that would also be present in this
comparison.

\begin{figure}
\centerline{\epsfig{file=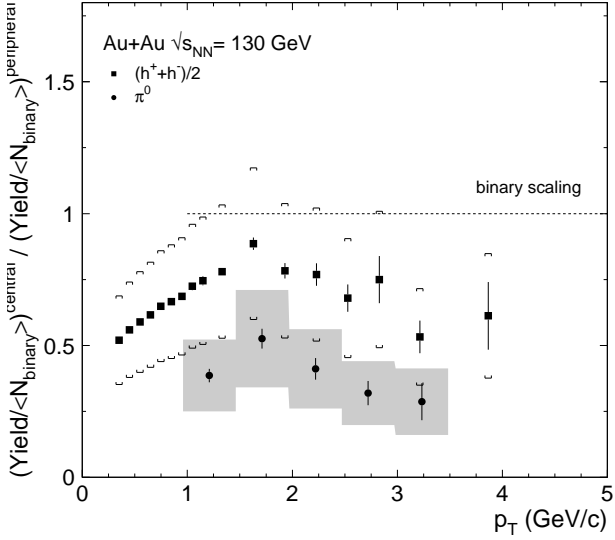,width=1.0\linewidth}}
\caption[]{
	Ratio of yield per event in central {\em vs} 
	peripheral Au+Au collisions, with each divided 
	by $\langle N_{binary} \rangle$ for that class.
	For $\pi^{0}$ the weighted average
	of PbSc and PbGl results is shown.
	The error bars indicate the statistical 
	errors on the spectra.  The surrounding bands 
	[shaded for $\pi^{0}$'s, brackets for 
	$(h^{+}+h^{-})/2$] are the 
	quadrature sums of (i) the parts of the systematic errors on 
	the spectra that do not cancel in the ratio,
	and (ii) the uncertainty in $\langle N_{binary} \rangle$
	(see Table~\ref{tab:syst}).
	}
\label{fig:ratio_central_periph}
\end{figure}

The central-to-peripheral ratios are plotted in 
Fig.~\ref{fig:ratio_central_periph}.  Like $R_{AA}$ this ratio is
below unity at all observed $p_T$ for both charged hadrons and neutral 
pions, indicating a suppression of the yield per N+N
collision in central collisions relative to peripheral.
The difference between the two ratios 
implies that the $\pi/h$ ratio is smaller in central collisions 
than in peripheral.


%

We have presented spectra for charged hadrons and 
neutral pions measured at 90$^o$ from central and 
peripheral Au+Au 
collisions in the PHENIX experiment at RHIC.  
Above $p_T\sim 2$~GeV/c, 
the spectra from peripheral collisions appear
to be consistent (albeit within a substantial systematic
error) with a simple, incoherent sum of underlying
N+N collisions.  The spectra from central
collisions, in contrast, are systematically below the
scaled N+N expectation, both when compared
to data from p+p collisions and to spectra from Au+Au 
peripheral collisions. 
The suppression in central collisions 
is in qualitative agreement with the predictions of energy 
loss by scattered partons traversing a dense medium.
However, other nuclear medium effects should be understood
before a quantitative conclusion can be drawn.  Measurements
in p+A at RHIC can help in this direction.

%


We thank the staff of the RHIC project, Collider-Accelerator, and Physics
Departments at BNL and the staff of PHENIX participating institutions for
their vital contributions.  We acknowledge support from the Department of
Energy and NSF (U.S.A.), Monbu-sho and STA (Japan), RAS, RMAE, and RMS
(Russia), BMBF, DAAD, and AvH (Germany), FRN, NFR, and the Wallenberg
Foundation (Sweden), MIST and NSERC (Canada), CNPq and FAPESP (Brazil),
IN2P3/CNRS (France), DAE and DST (India), LG-YF, KRF and KOSEF (Korea),
and the US-Israel Binational Science Foundation.



\vspace{-1cm}

\begin{table}
\caption[]{Relative systematic errors on hadron yields and 
	central-to-peripheral ratios.
	The errors are quoted for representative $p_{T}$ and 
	vary between the values shown.
	For the charged hadron ({\it h}) 
	data the errors are highly correlated in $p_{T}$ for both 
	yields and ratios.
	For the $\pi^0$ data, approximately half of the error in 
	the yield is perfectly correlated in $p_{T}$, and some 
	correlation remains in the ratio. 
	}
\begin{tabular}[]{lccccc}
Sys. error: & Yield & $p_{T}$ [GeV/c] & & Cent/Per  & $p_{T}$ [GeV/c] \\ 
\hline
{\it h} data & 27\% 	& 0.5 & 		& 8\% & all\\
	     & 16-18\% 	& 0.8-3.5	& &	&\\
	     & 30\%	& 4.7		& &	&\\ 
\hline 
$\pi^0$ data & 25\%	& 1.2	&	& 24\% & 1.2 \\
(PbSc)		& 35\%	& 3.7	&	& 33\%	&  3.2 \\ 
\hline
$\pi^0$ data & 33\%	& 1.2	&	& 32\% & 1.2 \\
(PbGl)		& 52\%	& 3.7	&	& 40\%	& 2.7 \\ 
\hline
$\pi^0$ data & 21\% & 1.2        &   & 20\% & 1.2 \\
(combined)       & 30\% & 3.7     &      & 24\% & 2.7 \\ 
\hline
N+N ref.	& 20\%  & 1.0	&	& N/A &\\
		& 35\%  & 5.0	&	&     &\\ 
\hline
$\langle N_{binary} \rangle$ & & & & 29\% & all\\
central & 11\%	& all	& & & \\
peripheral	& 30\%  & all    && 	& \\ 
\end{tabular}
\label{tab:syst}
\end{table}

\end{multicols}


\begin{references}
\bibitem{foot2}The error due to model uncertainties is 2\%~\cite{mult}. 
An additional 3.5\% error
results from time dependencies in the centrality selection over the 
large data sample.
\end{references}
\end{document}